\documentclass[12pt]{article}
\catcode`\@=11
\@addtoreset{equation}{section}

\def\@normalsize{\@setsize\normalsize{15pt}\xiipt\@xiipt
\abovedisplayskip 14pt plus3pt minus3pt%
\belowdisplayskip \abovedisplayskip
\abovedisplayshortskip \z@ plus3pt%
\belowdisplayshortskip 7pt plus3.5pt minus0pt}

\def\small{\@setsize\small{13.6pt}\xipt\@xipt
\abovedisplayskip 13pt plus3pt minus3pt%
\belowdisplayskip \abovedisplayskip
\abovedisplayshortskip \z@ plus3pt%
\belowdisplayshortskip 7pt plus3.5pt minus0pt
\def\@listi{\parsep 4.5pt plus 2pt minus 1pt
     \itemsep \parsep
     \topsep 9pt plus 3pt minus 3pt}}

\relax
\catcode`@=12
\catcode`\@=11
\def\section{\@startsection{section}{1}{\z@}{3.5ex plus 1ex minus
   .2ex}{2.3ex plus .2ex}{\large\bf}}
\def\thesection{\arabic{section}.}

\def\appendix{\setcounter{section}{0}
 \def\thesection{Appendix \Alph{section}:}
 \def\theequation{\Alph{section}.\arabic{equation}}}

\begin{document}

\begin{titlepage}
\begin{center}
{\Large   Confinement,     Chiral Symmetry Breaking \\
and   Faddeev-Niemi Decomposition  \\
in  QCD

 }
\end{center}

\vspace{1em}
\begin{center}
{\large   Kenichi Konishi$^{(1,2,3)}$ and  Kazunori Takenaga$^{(2,1   *)}$}

\end{center}
\vspace{1em}
\begin{center}
{\it
Dipartimento di Fisica -- Universit\`a di Pisa $^{(1)}$\\
Istituto Nazionale di Fisica Nucleare -- Sezione di Pisa $^{(2)}$\\
Via Buonarroti, 2, Ed.B-- 56127 Pisa (Italy)  \\
Department of Physics -- University of Washington $^{(3  **)}$ \\
Seattle, WA 19185 (USA)  \\
E-mail:   konishi@phys.washington.edu,   takenaga@df.unipi.it; \\
 konishi@df.unipi.it;

}
\end{center}
\vspace{3em}
\noindent
{\bf ABSTRACT:}
{   We  identify   two distinct, complementary classes of  gauge field configurations
for QCD  with  $SU(2)$ gauge group,   one  ({\it instanton-like
configurations}) having to do with chiral
symmetry breaking but not  with confinement,  the other ({\it regularized
Wu-Yang monopoles})
very    likely responsible for confinement but unrelated to chiral symmetry
breaking.
Our argument is based on a semiclassical analysis of fermion zero modes in
these backgrounds,
made by use of a gauge field decomposition recently introduced by Faddeev
and Niemi.  
Our result suggests that the two  principal  dynamical phenomena in QCD,
confinement and chiral
symmetry breaking,  are   distinct effects,   caused by two
 competing classes   of gauge field configurations.  
  }

\vspace{1em}

\begin{flushright}
November   1999\\
IFUP-TH 61/99
\end{flushright}

\bigskip
\begin{flushleft}

{*}) Address after   November 1999; Niels Bohr Institute, Denmark \\
{**}) Visiting  scholar (till June/2000).
\end{flushleft}

\end{titlepage}

\newcommand{\1}{{\Bbb I}}
\newcommand{\Z}{{\Bbb Z}}
\newcommand{\beq}{\begin{equation}}
\newcommand{\eeq}{\end{equation}}
\newcommand{\bea}{\begin{eqnarray}}
\newcommand{\eea}{\end{eqnarray}}
\newcommand{\beas}{\begin{eqnarray*}}
\newcommand{\eeas}{\end{eqnarray*}}
\newcommand{\defi}{\stackrel{\rm def}{=}}
\newcommand{\non}{\nonumber}
\def\dirac{{\cal D}}
\def\dplus{{\cal D_{+}}}
\def\dminus{{\cal D_{-}}}
\def\dbar{\bar{D}}
\def\L{{\mathcal L}}
\def\H{\cal{H}}
\def\de{\partial}
\def\si{\sigma}
\def\sb{{\bar \sigma}}
\def\rn{{\bf R}^n}
\def\r4{{\bf R}^4}
\def\s4{{\bf S}^4}
\def\ker{\hbox{\rm ker}}
\def\dim{\hbox{\rm dim}}
\def\sup{\hbox{\rm sup}}
\def\inf{\hbox{\rm inf}}
\def\infi{\infty}
\def\nrm{\parallel}
\def\nrmi{\parallel_\infty}
\def\om{\Omega}
\def\Tr{ \hbox{\rm Tr}}
\def\const{\hbox {\rm const.}}
\def\o{\over}
\def\th{\theta}
\def\im{\hbox{\rm Im}}
\def\re{\hbox{\rm Re}}
\def\bra{\langle}
\def\ket{\rangle}
\def\Arg{\hbox {\rm Arg}}
\def\Re{\hbox {\rm Re}}
\def\Im{\hbox {\rm Im}}
\def\diag{\hbox{\rm diag}}

It is now  widely accepted that  confinement in Quantum Chromodynamics (QCD)
is a  kind of dual superconductivity \cite{Nam,Thoo,DG}.    According to
this  idea,  the ground state of
QCD  is a condensate of magnetic monopoles.  Color electric fields are
expelled from the hadronic
medium,  except  as  possible thin filaments (Abrikosov votex) connecting
quarks
and antiquarks,   leading to a linearly
rising potential  between them.

The details of this phenomen is difficult to analyse, though,    due to the
fact that
the role of   the Cooper pairs  in the standard superconductors   is here
played by  topologically
nontrivial  soliton-like configurations.   Actually, the situation is
worse   since,  unlike  what happens in
theories   with elementary   scalar fields in the adjoint representation,
magnetic monopoles  in QCD  do
not correspond to any  stable   solitonic    solutions  of the  classical
Yang-Mills equations of
motion.   In fact such a static solution is known not to exist in pure
Yang-Mills theories.
 Rather  they  represent   sets  of   regularized Wu-Yang monopole
configurations,    whose  presence
may be conveniently  detected by  't Hooft's  Abelianization
procedure \cite{Thoo}.

A  new   decomposition  of  the Non-Abelian gauge fields recently  proposed
by
Faddeev and Niemi \cite{fn,FNN}      makes     these ideas more concrete;
it
appears to enables us   to analyse    the relevant  physics aspects in more
detail,
without the need  of  choosing
particular gauges such as the maximally Abelian gauge \cite{MaxA}.

   As an
example,     we discuss here    the possible   connection between
confinement and
chiral symmetry breaking in QCD.
We shall argue that  configurations responsible for confinement
and for chiral symmetry breaking
are quite distinct  and in a sense complementary,   in QCD.

 A key aspect in discussing such  a connection  is the   existence of the
fermion zero modes
in the background of the semiclassical  monopoles \cite{JR}.  In fact,
the connection between these two dynamical phenomena
  has been recently    illucidated
in full nonperturbative  analysis,   in many
$SU(n_c)$ or $USp(2n_c)$   theories    with $N=2$ supersymmetry (broken
softly to $N=1$) \cite{SW,CKM}.
We shall not discuss here physics of these models (which show  very rich
varieties   of
nonperturbative scenarios, some resembling  those in the actual world of
strong interactions, some quite
different);  for the present purpose   we shall only  draw one  lesson from these studies.
Namely   the existence of the fermion zero
modes in the background of semiclassical  monopoles,    is a necessary
condition for the  low-energy,  light
magnetic monopoles to carry flavor quantum  numbers,  hence  a condition
for  their  condensation
(confinement) to imply chiral symmetry breaking.

The models of \cite{SW,CKM}    contain   't Hooft-Polyakov
monopoles \cite{TM}   in the spectrum.  The
existence and number of the zero modes for
each fermion can be (and has been) established  both explicitly (by
generalizing the analysis by Jackiw and
Rebbi)   and through Callias'  index theorem \cite{indCa}.   The  resulting
semiclassical  flavor multiplet  structure
can be  compared with  the spectrum of  low-energy massless (fully quantum
mechanical)  monopoles,  with
complete   matching between the two \cite{CKM}.

In  QCD,  one might instead   use   the background configurations  which
are presumably   dominant
in its  ground state \cite{Nambu,CDG}.    For simplicity, we discuss here 
QCD with $SU(2)$ gauge group.
Candidate configurations we shall consider are  (regularized) Wu-Yang
monopoles,  instantons, and
modification/collection of these.       According to Faddeev and Niemi,
the   $SU(2)$
connection  can be decomposed  as
 \beq A_{\mu}^a= C_{\mu} {\bf n}^a + {\tilde \sigma}(x) ( \de_{\mu} {\bf
n}\times
{\bf n})^a + \rho \, \de_{\mu} {\bf n}^a; \qquad   {\tilde \sigma}(x) = 1 +
\sigma(x),    \label{FadNie} \eeq
in terms of  the unit vector field ${\bf n}$  and   the Abelian  gauge
field $C_{\mu}$
which live  on $S^2$  and $S^1$ factors,  respectively, of $SU(2)
=S^3 \sim  S^2 \times S^1$ manifold,  and  a charged "scalar"  field
\beq    \phi= \rho(x)+ i \sigma(x).\eeq
 In  terms of these variables,
the Wu-Yang  singular monopole solution \cite{Wuyang},  for instance,  is
\beq     n^a=   { x^a \o r}, \quad   C_{\mu}= \phi=0.  \eeq
Thanks to the presence of the other degrees of freedom,   one might think
that   in the
dominant configurations   such singularities are
 actually     regularized by the zero of
$1-|\phi|^2$.

   Indeed,   the standard Yang-Mills
action written   in these variables reads
\bea   S   &=& -{ 1\o 4  g^2}   \int d^4 x  \, F_{\mu \nu}^2  =   -{ 1\o 4
g^2}   \int d^4 x  \, \{  \,    [G_{\mu \nu } +
(1-|\phi|^2)
\, H_{\mu {\nu}}]^2  +  \non    \\ &+& 2 \, [\sum_{\mu \ne {\nu}}
\de_{\mu} {\bf n} \cdot \de^{\mu}  {\bf n}  \,
(D_{\nu} \phi)^{*} (D^{\nu} \phi)    -
\sum_{{\mu} \ne {\nu}}  \de_{\mu} {\bf n} \cdot \de_{\nu}  {\bf n}  \,
(D^{\mu} \phi)^{*} (D^{\nu} \phi )]  \non \\
&- &   i \, [  (D^{\mu} \phi)^{*} (D^{\nu} \phi) -  (D^{\nu}\phi)^{*}
(D^{\mu} \phi) ] \, H_{{\mu}{\nu}}
\,  \} +  \{\theta \,\, {\hbox {\rm - term}}\},
\label{FNaction} \eea
\newpage 
\noindent  where  \footnote{The $\theta$   term has also quite an elegant form,
\bea  { \theta \o 32 \pi^2}  \int d^4 x  \, F_{\mu \nu}  {\tilde F}^{\mu
\nu}   &=&    { \theta \o 64 \pi^2}
\epsilon_{\mu
\nu \rho   \sigma}   \int d^4 x  \,
 \{  [G^{\mu \nu } + (1-|\phi|^2)  \, H^{\mu {\nu}}] [G^{\rho  \sigma } +
(1-|\phi|^2)  \, H^{\rho \sigma}]
\non \\ &+&  i  \,[  (D^{\mu} \phi)^{*} (D^{\nu} \phi) -  (D^{\nu}\phi)^{*}
(D^{\mu} \phi) ] \,
H^{{\rho }{\sigma}} \}. \non \eea}
\beq  G_{{\mu}{\nu}}=\de_{\mu} C_{\nu} - \de_{\nu} C_{\mu}; \quad  \phi=
\rho+ i \sigma, \quad
H_{{\mu}{\nu}}=  ({\bf n} \cdot  \de_{\mu} {\bf n} \times \de_{\nu} {\bf
n});\eeq
\beq   D_{\mu} \phi =   (\de_{\mu} - i  C_{\mu}) \phi. \eeq
In the case of the Wu-Yang monopole,   the singularity in the energy   at
the origin comes from the behavior
$  H_{i j}^2  =  2/r^4.$

 Note that the local $U(1)$ invariance  (corresponding to $SU(2)$ gauge
transformations
of the form  $U=  \exp{i \alpha   {\bf n}
\cdot  {\bf \tau}/2}$)       is manifest in Eq.(\ref{FNaction}).   Fixing
the direction of ${\bf n} $  (by gauge
transformations  belonging to    $SU(2)/U(1)$)   as   ${\bf n}= (0,0,1)$,  for example,   amounts   to the Abelian
gauge fixing:    for the ${\bf n}$
configuration of the form, $ n^a=   { x^a \o r}$,  such a gauge
transformation
 would  introduce    an  (apparently) singular
Dirac monopole,  even though the gauge field configuration itself  is perfectly
regular.

A curious feature of Faddeev-Niemi decomposition,  that the connection
contains ${\tilde \sigma}(x)
=1 + \sigma(x) $ (and  $\rho(x)$)    naturally   while   the action depends
on   $\sigma(x) $ (and $\rho(x)$)
in a simple  and  significant way,  is central  to  our
discussion.

Before going into our  main argument,  let us note, following Faddeev and
Niemi,  that the form of
Eq.(\ref{FNaction})  suggests a very clear picture of different possible
phases of QCD.  Namely,   if
the  field   $\phi(x)$ fluctuates  more strongly   than  the   ${\bf n}$ field,
one  could integrate   the former first,
in the sence of renormalization group,   arriving at a low-energy effective
action of QCD, 
\beq    S^{eff}_{FN}  =  \int d^4x \,   \{   \Lambda^2 \sum_{\mu}
\de_{\mu} {\bf n} \cdot
\de_{\mu}  {\bf n}   +  \sum_{\mu \ne {\nu}} ({\bf n} \cdot  \de_{\mu} {\bf
n} \times \de_{\nu} {\bf n})
^2 \}, \eeq
suggested in \cite{fn}, describing  the confinement  phase of QCD  ($\Lambda$  is a dynamically
generaled mass).
This action  has two important features, one being the   unique action
containing  ${\bf n}$
field and   allowing for Hamiltonian  interpretation,  and second,
containing   topological solitons
\cite{fnnat}   which could be thought as models of gluonia.  Though very
interesting,  it  is not our main interest here  to  pursue     these ideas
further.

Vice versa,   if   dynamically    ${\bf n}$ field fluctuates  more,
renormalization group flow would
instead   yield   a low energy  action which looks more like
\bea   S^{eff{'}}_{FN}    &=& - \int d^4 x  \, \{  \,    G_{\mu \nu }^2
-   (D_{\nu} \phi)^{*} (D_{\nu} \phi)
  +    (1-|\phi|^2)^2  \}
\label{FNactionbis } \eea
 which is   the standard  Higgs model (describing a possible Higgs phase).
Finally,   if   none of the fields
fluctuate strongly, then one would have  the original  Yang-Mills action
at   low  energies: one is in the
non-Abelian Coulomb phase in this case (this will be the case if
a sufficient number of massless fermions are
present).

Though  obviously  over-simplified,   these observations do  show that
field configurations which
contribute   to nonvanishing vev  $\bra \phi \ket  \ne 0 $  tend  to bring
the system to the Higgs phase
(short-ranged color electric force);      confinement  is caused by
configurations giving rise to  dual Higgs
phase (hence with
 $\bra \phi \ket  =  0 $).

A zero-energy static  left-handed  quark  field  satisfies  the equation
(we shall consider the massless quarks
for simplicity)
\beq  i (\sigma_k )_{\alpha \beta}  [ \de_k - i (C_k n^a +   {\tilde
\sigma} ( \de_k   {\bf n}\times {\bf n})^a + \rho
\, \de_{k} {\bf n}^a ) {\tau^a \o 2}]_{ij }  \, \psi_{L \beta j}=0.
\label{equation} \eeq
which follows from    the standard covariant    Weyl equation
\beq    i {\bar \sigma}_{\mu}  (\de_{\mu}  - i  A_{\mu} ) \psi_L=0, \eeq
with  $A_{\mu}$  taken in the Faddeev-Niemi form.  The exact $U(1)$  local
invariance was    used to set
$C_0(x)=0$ above.

As is well known the Dirac Hamiltonian (the operator multiplying $\psi_L$
in  Eq.(\ref{equation}))  commutes
with  the total  "angular momentum" operator,
\beq  {\bf J}=  {\bf L}  +{\bf s}  +   {{ \bf \tau }\o 2 },\eeq
composed of the orbital and spin angular momenta  {\it and}  the gauge
$SU(2)$ spin.  For this reason one seeks
for  a singlet    zero mode having the general structure
\beq  (\psi_{L })_{\alpha i}= -i \tau^2_{\alpha i} \,g_L(r) + (\tau^a
\tau^2)_{\alpha i} \,r^a h_L(r).
\label{general}\eeq
The functions    $g_L(r)$ and $ h_L(r)$ satisfy the coupled equations,
($  j_L(r)   \equiv   r h_L(r)$):
\bea   j_L^{'} + {2 \o r} j_L -    {   {\tilde \sigma} \o  r}   j_L   +  [{
r^2   C(r) \o 2
}      + { \rho \o r}  ] \,  g_L &=&  0;
\non \\    g_L^{'}     +   {  {\tilde \sigma}
\o  r}  g_L  -     [  {  r^2   C(r)   \o 2}  \,   -  {  \rho \o  r  }    ] \,
j_L &=& 0,
\label{zeroeq} \eea
where we  have  set
\beq    C_k(x)= r^k C(r), \eeq
and assumed spherically symmetric  forms for $C(r),\, \rho(r),\, $ and $\sigma(r)$.

Let us first consider  the case of an instanton background at $x_0=0$,
\beq
n^a(x)=  {x^a \o r}; \qquad {\tilde \sigma}(r)   =  { 2 } \, r^2 \,  f(x);
\qquad f(x)=  {1 \o r^2+ \lambda^2};   \qquad
C(r)=\rho(r)  =0,
\label{instanton}\eeq where $\lambda$ is the instanton size.
This is actually only the instanton
configuration at
$x_0=0$,  the static configuration relevant to the {\it three } dimensional
zero mode. We shall loosely refer to it
as the  instanton below. Note that  the monopole  singularity at the origin
is smoothened by the zero of
$1-|\phi|^2$     since  ${\tilde \sigma} (x) \to   0$  (or   $\sigma(x) \to
-1$)   as $r \to 0$  and
$\rho\equiv 0$.    Eq.(\ref{zeroeq})    can be immediately
integrated in this case and gives
\beq     g_L=  e^{- \int_0^r dr  \,   {  {\tilde \sigma} \o r}}=   {1 \o
r^2 + \lambda^2};    \qquad    j_L=0,
\label{instzd}
\eeq  which is the well-known three-dimensional  zero mode of the
lefthanded fermion \cite{Kiskis},  related
to  the  four-dimensional   Euclidean zero mode  by the spectral-flow
argument \cite{CDG}.  The instanton background
at
$x_0\ne 0$  does not allow for three-dimensional fermion zero modes.

Note also that  there is   another independent solution of  Eq.(\ref{zeroeq})
\beq  g_L=0;  \quad    j_L(r) =   \exp{- \int_0^r  dr  { 2-  {\tilde
\sigma} \o r}},
\label{nonnorm} \eeq
which is however not  normalizable.

Of course, the precise form of the instanton background is not needed for
the
existence of  the normalizable   fermion zero mode.
The essential features of the background  (\ref{instanton})  are:
\begin{description}
\item[i)]  the hedgehog form of the ${\bf n}$ field;
\item[ii)]  the behavior of the   $\sigma(r)$  field (in the gauge $rho=0$),
\beq   {\tilde{\sigma}}(0)=0; \qquad  {\tilde{\sigma}}(r)   \stackrel   {r
\to \infty}{\longrightarrow} 2
\quad ({\hbox {\rm or}}  \quad     {\sigma}(r)   \stackrel   {r \to
\infty}{\longrightarrow} 1).
\label{conditions}  \eeq
\end{description}
In fact, the first of (\ref{conditions})  guarantees that the gauge
configuration is regular at the origin
and that   the zero mode is normalizable at the origin,  while   the
second, which means that  it looks like a
monopole of charge
$2$  from the distance,   leads to   the asymptotic behavior of the fermion
zero mode,
$g_L \sim 1/r^2$,  compatible with the normalizability at $r \to
\infty$.

Another  example of  configuration of this class  is the di-meron configuration \cite{Fubini, Kiskis}
\bea
n^a(x) &=&  {x^a \o r}; \qquad {\tilde \sigma}(r)   =  r {\de \o \de r}  R(x); \quad C(r)=\rho(r)  =0, \non \\
R(x)&= &\log [r^2 + y^2 - t^2 + \{ r^2 + (t-y)^2\}^{1/2} \{ r^2 +(t+y)^2\}^{1/2}];  \qquad
\eea 
for  $-y < t < y$ ($2y$ is the two-meron separation).  

In fact we can  consider  a more general  class  of configurations with
these characteristics,  with
possibly
$C(r)  \ne 0$, $\rho(r) \ne 0$,    and call  these  {\it  instanton-like configurations}.
Although   one has  coupled equations for the system $(g_L,
j_L)$,
one can quite generally assume that  a    normalizable solution  $(g_L, j_L)$
exists if $C(r)$ and $\rho$ are sufficiently small.

The instanton-like configurations may  then  well have to do with the
chiral symmetry breaking,
since  in  the collection of such configurations  (``instanton liquid" \cite{Dyakonov})  the
chiral
symmetry breaking vevs
\beq  \bra {\bar u_R} u_L \ket =
\bra {\bar d_R} d_L \ket  \ne 0 \label{XSB}\eeq
will   be  nonvanishing. 

Instanton-like configurations are on the  other hand  of  no use from the
point of view of confinement.
In fact,  the asymptotic behavior of the $\sigma$ field means that $|\phi|
\to 1$ at infinity,
suggesting   that   instanton-like configurations tend to bring the system into
the Higgs phase \footnote{See \cite{Calorons} for recent related ideas.}.  
This is  consistent with the general idea that the instantons, being
point-like in four dimensions,
have nothing
to do with confinement,  but our argument is based on the three dimensional  properties
of these configurations. 

Consider next the class of configurations  with the following characteristics:
\begin{description}
\item[i)]   the ${\bf n}$ field is the  hedgehog form $ n^a = x^a/r$;
\item[ii)]   the   $\sigma(r)$  field behaves as (we set $\rho=0$ by the $U(1)$ gauge transformation),
\beq   {\tilde{\sigma}}(0)=0; \qquad  {\tilde{\sigma}}(r)   \stackrel   {r
\to \infty}{\longrightarrow}
1, \label{conditionsbis} \eeq
\end{description}
namely
\beq  A_{i}^a=  {\tilde \sigma}(x)  \, \epsilon_{aij}  { x^j \o r^2} +
\ldots   \label{asym}  \eeq
which we call   {\it  regularized  Wu-Yang monopole}     configurations.
The difference of factor $2$ in the asymptotic behavior of $ {\tilde
\sigma}(x)$ field
as compared to the instanton-like configurations,  is crucial.  It means
that, on the one hand,
$\sigma(r) \to 0$ ($\phi \to 0$)  at infinity  so   these configurations
are consistent with confinement;  on the other hand, it implies that the fermion zero modes are 
non-normalizable (see below).  

Note that  the single meron configuration at a fixed time slice
\bea
n^a(x) &=&  {x^a \o r}; \qquad {\tilde \sigma}(r)   =  r {\de \o \de r}  R(x); \quad C(r)=\rho(r)  =0, \non \\
R(x)&= &\log [\{r^2 + t^2\}^{1/2}   -t  ], \qquad
\eea 
is of this type: in fact this class of configurations may alternatively be   called meron-like configurations.

If
(collection of) these configurations are indeed dominant in the ground
state of QCD,     it
will lead to dual Meissner    effect (confinement), as suggested by 't
Hooft, Mandelstam and Nambu.

These configurations however    do not trigger   chiral symmetry breaking,
since  there are  no normalizable fermion zero
modes in this case.   In fact if we assume   $C(r)=0$   for simplicity,
one of the solutions of the
zero-energy equation is again
   \beq   g_L=  e^{- \int_0^r   dr  \,   {  {\tilde \sigma} \o r}}  \eeq
but this is not normalizable at infinity,   because it behaves as $  \,g_L
\sim 1/r$.    In the case of the meron configuration this solution   coincides with 
  the one given in \cite{Kiskis}.     The other
solution
$j_L(r)$ (Eq.(\ref{nonnorm}))    is non-normalizable in this case,   too.    Again, we could allow
for nonvanishing $C(r)$ and $\rho(r)$    fields, but it
is clear that  for quite general class of perturbations these solutions
will remain both non-normalizable.

We conclude that regularized Wu-Yang  monopoles (or rather collections of
those)
are fundamental for confinement but have  in themselves   nothing to do with chiral
symmetry breaking.
It is quite remarkable that one can identify,  through the analysis of
semiclassical fermion zero
modes,     two distinct and  complementary  sets of configurations:  one
({\it instanton-like
configurations}) has  likely to do with chiral symmetry breaking but with
no relation to confinement and  the
other  ({\it regularized Wu-Yang monopoles})   being  most likely
responsible for confinement but are,
as they are,    unrelated  to  chiral symmetry breaking.
This suggests  that  the two main nonperturbative effects of  QCD, confinement
and chiral symmetry
breaking,   are distinct effects,  and in particuar that    the latter
(chiral symmetry breaking) is
not a direct consequence of the former  (confinement).
 There are hints that support this conclusion   in the lattice approach to
QCD \cite{Lat}.

In fact, our conclusion overlaps considerably with that of Callan et. al. \cite{CDG}, 
but the use of the Faddeev-Niemi
decomposition appears to allow for a particularly  simple way  to relate  
the question of chiral symmetry breaking  
(existence or absence of the fermion zero modes) to that of   confinement.

We conclude with several comments.
\begin{enumerate}

\item   The fact that the asymptotic value of $ \sigma$   is quantized (in
the gauge $\rho(x)=0$),
\beq   {{\sigma}}(r)   \stackrel   {r \to \infty}{\longrightarrow}
n, \quad n \in {\bf Z}, \label{quant} \eeq
is   fundamental to  our discussion.   (\ref{quant})  must be imposed  on
the  Faddeev-Niemi construction.    This
is so  because   in the gauge
${\bf n}=(0,0,1)$ and for $\rho=0$,  
$\tilde \sigma(\infty)$ represents the charge of the magnetic monopole (see
Eq.(\ref{asym})),  which  should
 obey
 Dirac's  quantization condition for   consistency if
fermions  are present in the theory.   It should be noted  that
although   ${\bf n}$ field   represents $\Pi_2(SU(2)/U(1))$  hence  
can be devided  into integer classes
of  winding number $S^2 \to S^2$, this
fact alone   is not sufficient to guarantee  the integer monopole charge.

\item   Our analysis seems to shed   some new  light on   the long-standing
question of the  interplay  between
the instantons and the monopoles in QCD.    Although they are  clearly not
totally unrelated,     their relations 
\cite{InstMon}   can be rather subtle.  
For instance,    in  the exact  Seiberg-Witten solution \cite{SW} of $N=2$ supersymmetric gauge
theories,     the 
renormalization of the $\theta$ parameter    is    due to
the  infinite sum of instanton
contributions,     but in the dual language more adequate at low energies  it
is seen  as due  to  the perturbative
loops of magnetic monopoles.   See also an earlier related idea \cite{Rossi}  and recent related 
results  in finite temperature QCD \cite{Calorons}.

\item  It is known that confinement can occur  without  chiral symmetry breaking,    
  as   is exemplified in many   supersymmetric models, supporting our idea that 
these two phenomena are in principle   distinct.    For example, in  
the  massless $N=1$ supersymmetric QCD with $n_f=n_c+1,$   the   low-energy degrees of freedom are 
mesons and baryons and their superpartners (confinement),  while chiral symmetries remain
unbroken in one of the possible vacua \cite{ADSAK}. 
It is suggestive that this occurs precisely in a theory in which
the effects  of instantons are known to be relatively  weak \cite{ADSAK}.

\item  The work of Faddeev and Niemi generalizes   an earlier
proposal   by Cho \cite{Cho},
who   required  that the gauge connection satisfy
\beq    {D}_{\mu}  {\bf n}  =0,  \eeq
whose solution is simply
\beq   A^a_{\mu}(x) = C_{\mu}  n^a + ( \de_{\mu} {\bf n}\times
{\bf n})^a. \eeq
The field tensor decomposes  as a sum of ``electric" and ``magnetic" parts,
\beq F_{\mu \nu}=  G_{{\mu}{\nu}}+ H_{{\mu}{\nu}}; \quad
G_{{\mu}{\nu}}=\de_{\mu} C_{\nu} - \de_{\nu}
C_{\mu};  \quad  H_{{\mu}{\nu}}=  ({\bf n} \cdot  \de_{\mu} {\bf n} \times
\de_{\nu} {\bf n}),\eeq
where   $ H_{{\mu}{\nu}}$  contains the (Wu-Yang) magnetic monopole  for
${\bf n}$  field of hedgehog
form. The decomposition by Faddeev and Niemi continues to share this nice
property, but having extra
degrees of freedom  ($\phi(x)$)   is capable of  accomodating   larger
classes of regular field
configurations  which are needed in our discussion.

\item Actually, the decomposition of  the $SU(2)$ connection  used here is
not a fully general one.
$A^a_{\mu}(x)$ is expressed in terms of six     independent functions
(after taking into account of the
local $U(1)$ invariance, one has  two   physical degrees of freedom in
$C_{\mu}$, two for ${\bf n}$ and
two more from $\phi(x)$),  which    matches   the number of the physical
degrees of freedom of
$A^a_{\mu}(x)$.      They were shown to be complete \cite{fn} in the sense
that  Yang-Mills equations of
motions  are  correctly reproduced by those for  
$C_{\mu}(x),{\bf n}(x), \phi(x)$.
   Of course,  as functional  integration variables  one needs  two more
degrees of freedom,  and the
  Faddeev-Niemi  decomposition has been  accordingly    generalized
recently \cite{FNOFF}.
The original  ``on-shell"  decomposition, however, seems to  contain just
sufficient number
of degrees of freedom (i.e., a  just   sufficiently  wide   classes of
configurations)  for the purpose of the
present exploratory  study.

\item  Very recently  Davies et.al.\cite{DHKM}   have computed the gluino
condensate in   pure $N=1$  supersymmetric
Yang-Mills theory,  by compactifying the time  coordinate on a cylinder,
and by using the gluino zero mode in
the background of the standard  BPS monopole for $A^a_i(x)$  ($A^a_0(x)$
appears as the  Higgs scalar).
One might wonder whether the absence of  fermion zero modes in the background
of the    regularized Wu-Yang monopole  we noted, is compatible with  their
work.   In fact,   if one repeats our
analysis  in the case of a fermion in the adjoint representation of the
gauge group,   one does find
a   pair of  zero modes  in the background   of  regularized Wu-Yang
monopole background.  The simplest way
to get   this result is to  set  the Yukawa coupling to zero  in the
original   analysis of Jackiw and Rebbi \cite{JR}.
In the case of the fundamental  fermion, it can be seen that the
normalizable Jackiw-Rebbi zero mode
becomes  non-normalizable \footnote{One of the authors (KK)  thanks H.
Terao for a discussion on this point.},  which is  
 compatible with  the absence of normalizable zero modes we found here.
  On the   other hand,   the
(pair of) normalizable zero modes found in
\cite{JR}   continue to be normalizable in the limit of the  vanishing Yukawa
coupling   in the case of an  adjoint  fermion \footnote{The asymptotic behavior of the zero modes
does change however in this limit: both zero modes now  behave as $1/r^2$,  in agreement    with the result of
\cite{DHKM}. }  
which is  consistent    with  the calculation of  \cite{DHKM}. 

\item    In
an alternative   QCD  with quarks in the adjoint
representation,   only  the regularized Wu-Yang configurations would lead to confinement
 while chiral symmetry breaking can be caused by both types of configurations. 
It is    thus possible that  at  some finite temperature chiral symmetry breaking can occur 
without confinement,   as suggested 
 by Kogut et. al.  \cite{Lat},    if   instanton-like configurations dominate over  Wu-Yang monopole like configurations.  

\end{enumerate}

\bigskip
\noindent{\bf Ackowledgment}
The authors acknowledge useful discussions with Andrea Barresi,  Massimo  Campostrini,  Prem  Kumar  and Ettore  Vicari.
K.T. would like to thank the INFN, Sezione di Pisa,   for hospitality.


\begin{thebibliography}{99}
\bibitem{Nam} Y. Nambu, {Phys. Rev.} {\bf  D10}  (1974) 4262,  
 S. Mandelstam, {Phys. Lett.} {\bf 53B}  (1975) 476;
                             {Phys. Rep.}  {\bf 23C} (1976) 245 ;
                             {Phys. Rev.}  {\bf D19} (1978)  2391
\bibitem{Thoo} G. 't Hooft,  {Nucl. Phys.} {\bf B190} (1981) 455
\bibitem{DG}  L. Del Debbio, A. Di Giacomo and G. Paffuti,  {Nucl. Phys. B} (Proc. Suppl.) {\bf   42}
(1995) 231; 
\bibitem{fn} L.D. Faddeev and  A.J. Niemi,  {Phys. Rev. Lett.}
{\bf 82}  (1999) 1624, hep-th 9807069
\bibitem{FNN} L.D. Faddeev and  A.J. Niemi, {Phys. Lett.} {\bf 449B}
(1999) 214,  hep-th 9812090 v3
\bibitem{MaxA} S. Kronfeld, G. Schierholz and U.-J. Wiese, {Nucl.
Phys.} {\bf B293}  (1987) 461
\bibitem{JR} R. Jackiw and   C. Rebbi, {Phys. Rev.} {\bf D13}
3398 (1976)
\bibitem{SW} N. Seiberg and E. Witten,  {Nucl. Phys.} {\bf  B426}
(1994) 19;  {Nucl. Phys.}
{\bf B431} (1994) 484
\bibitem{CKM}   K. Konishi and H, Terao,
{Nucl. phys.} {\bf B511} (1998) 264,  { hep-th/9707005},    
G. Carlino, K. Konishi and H. Murayama,  in preparation
\bibitem{TM} G. 't Hooft, {Nucl. Phys.} {\bf B79} (1974) 276,
 A.M. Polyakov, { JETP Lett.} {\bf 20}  (1974) 194
\bibitem{indCa} C. J. Callias, {Commun. Math. Phys.} {\bf 62} 213
(1978),  E. Weinberg, {Nucl. Phys.} {\bf B167} (1980) 500,  J. de Boer, K. Hori and Y. Oz, 
hep-th/9703100 v3 (1998)
 \bibitem{Nambu} Y. Nambu,   Caltech preprint, CALT-68-634 (1977)
\bibitem{CDG}  C.G. Callan, Jr.,  R. Dashen and D.J. Gross, {Phys. Rev.}  {\bf D17} (1978)  2717
\bibitem{Wuyang}  T.T. Wu and C.N. Yang,  in  ``Properties of Matter Under
Unusual Conditions", Ed. H. Mark
and  S. Fernbach, Interscience, New York, 1969
\bibitem{fnnat} L.D. Faddeev and  A.J. Niemi, {Nature} {\bf 387} 58
(1997),
L.D. Faddeev and  A.J. Niemi,  hep-th 9705176,
J. Gladikowski and  M. Hellmund, {Phys. Rev.} {\bf D56} 5194 (1997),
R.A. Battye and  P.M. Sutcliffe,  hep-th 9808129,
R.A. Battye and  P.M. Sutcliffe,  hep-th 9811077,
 H. Aratyn, L.A. Ferreira and  A.H. Zimerman,  hep-th/9902141
 H. Aratyn, L.A. Ferreira and  A.H. Zimerman,  hep-th 9905079
\bibitem{Kiskis} J. Kiskis,  {Phys. Rev.} {\bf D18} (1978) 3690
\bibitem{Fubini} V. De Alfaro, S. Fubini and G. Furlan, {Phys. Lett.} {\bf 65B}
(1977) 1631, 
\bibitem{Dyakonov}  D.I. Dyakonov and V. Yu. Petrov, {Nucl. Phys.}
{\bf B272}  (1986) 457,
 E.V. Shuryak,  ``The  QCD vacuum, hadrons and the superdense
 matter",   World Scientific (1988)
\bibitem{Lat} J. Kogut, J. Polonyi, H.W. Wyld and D.K. Sinclair, {Phys. Rev. Lett.} {\bf 54}
(1985) 1980,
F. Karsch and M. L\"utgemeier,  {Nucl. Phys.} (Proc.
Suppl.) {\bf 73}  (1999) 444,
and
references therein
\bibitem{InstMon}   M.N. Chernodub and F.V. Gubarev,   { JETP Lett.}
{\bf 62}
(1995) 100, hep-th/9506026 v2,  H. Suganuma, K. Itakura and H. Toki,
hep-th/9512141 v2,
R. Brower, K.N. Oginos and    C.-I. Tan,  {Phys. Rev.} {\bf D55}
(1997)  6313
\bibitem{Rossi}  P. Rossi,  {Nucl. Phys.} {\bf B149}  (1979) 170
\bibitem{Calorons}  T.C. Kraan and P. van Baal, {Phys. Lett.} {\bf 428B}
(1998) 268, hep-th/9805168,  K. Lee and C. Lu,  {Phys. Rev.} {\bf D58}
(1998) 025011,   hep-th/9802108
\bibitem{ADSAK} I. Affleck, M. Dine and N. Seiberg, {Nucl. Phys.} {\bf B256}  (1985) 557,
V. Novikov, M. Shifman, A. Vainshtein and V. Zakharov, {Nucl. Phys.} {\bf B260}  (1985) 157,
 D. Amati, K. Konishi, Y. Meurice, G.C. Rossi and G. Veneziano,  {Phys. Rep.} {\bf 162}  (1988) 169,  
N. Seiberg,   {Nucl. Phys.} {\bf B435}  (1995) 129, hep-th/9411149
 \bibitem{Cho} Y.M Cho, {Phys. Rev.} {\bf D21} 1080 (1980);  ibid.
{\bf D23} 2415 (1981),  Y.M. Cho and D.G. Pak,  hep-th/9906198
\bibitem{FNOFF} L.D. Faddeev and   A.J. Niemi,  hep-th 9907180,
S.V. Shabanov,  hep-th 9907182
\bibitem{DHKM}  N.M. Davies, T.J. Hollowood, V.V. Khoze and M.P. Mattis,
hep-th/9905015
(1999)

\end{thebibliography}
\end{document}